\documentclass{ws-procs975x65ns}

\begin{document}
\newcommand{\lsim}{\mathrel{\rlap{\raise -.3ex\hbox{${\scriptstyle\sim}$}}%
                   \raise .6ex\hbox{${\scriptstyle <}$}}}%
\newcommand{\gsim}{\mathrel{\rlap{\raise -.3ex\hbox{${\scriptstyle\sim}$}}%
                   \raise .6ex\hbox{${\scriptstyle >}$}}}%

\title{Impact of foregrounds on Cosmic Microwave Background maps}

\author{G. De Zotti}

\address{INAF-Osservatorio Astronomico di Padova, Vicolo Osservatorio 5,
I-35122 Padova, Italy\\E-mail: dezotti@pd.astro.it}

\author{C. Burigana}

\address{IASF/CNR, Sezione di Bologna, Via Gobetti 101, I-40129 Bologna, Italy\\
E-mail: burigana@bo.iasf.cnr.it}

\author{C. Baccigalupi and R. Ricci}

\address{SISSA, Via Beirut 4, I-34014 Trieste, Italy\\
E-mail: bacci,ricci@sissa.it}

%%%%%%%%%%%%%%%%%%%%%%%%%%%%%%%%%%%%%%%%%%%%%%%%%%%%%%%%%%%%%%
% You may repeat \author \address as often as necessary      %
%%%%%%%%%%%%%%%%%%%%%%%%%%%%%%%%%%%%%%%%%%%%%%%%%%%%%%%%%%%%%%

\maketitle

\abstracts{We discuss the possible impact of astrophysical
foregrounds on three recent exciting results of Cosmic Microwave
Background (CMB) experiments: the WMAP measurements of the
temperature-polarization ($TE$) correlation power spectrum, the
detection of CMB polarization fluctuations on degree scales by the
DASI experiment, and the excess power on arcminute scales reported
by the CBI and BIMA groups. A big contribution from the Galactic
synchrotron emission to the $TE$ power spectrum on large angular
scales is indeed expected, in the lower frequency WMAP channels,
based on current, albeit very uncertain, models; at higher
frequencies the rapid decrease of the synchrotron signal may be,
to some extent, compensated by polarized dust emission. Recent
measurements of polarization properties of extragalactic radio
sources at high radio frequency indicate that their contamination
of the CMB polarization on degree scales at 30 GHz is
substantially below the expected CMB $E$-mode amplitude. Adding
the synchrotron contribution, we estimate that the overall
foreground contamination of the signal detected by DASI may be
significant but not dominant. The excess power on arc-min scales
detected by the BIMA experiment may be due to galactic-scale
Sunyaev-Zeldovich effects, if the proto-galactic gas is heated to
its virial temperature and its cooling time is comparable to the
Hubble time at the epoch of galaxy formation. A substantial
contamination by radio sources of the signal reported by the CBI
group on scales somewhat larger than BIMA's cannot be easily ruled
out. }

\section{Introduction}\label{sect:intro}

For a lucky, at least from the cosmologist's point of view,
coincidence, the spectral energy distribution of most classes of
extragalactic sources has a minimum at millimeter wavelengths,
fortuitously close to the peak of the Cosmic Microwave Background
(CMB) spectrum. At high Galactic latitudes the minimum is a factor
of few$\times 10^{-6}$ below the peak CMB intensity;
correspondingly, foreground fluctuations can be below the $\mu$K
level, allowing a clean detection of CMB anisotropies.

This ``cosmological window'', extending roughly over the frequency
range 40--150 GHz, is crucial to exploit in depth the
extraordinary information content of the CMB: in fact, with
current technologies the limit is not set by instrumental
sensitivity but by the contamination by astrophysical foregrounds.

The situation is actually slightly different when we are dealing
with CMB anisotropies rather than with total intensity. On angular
scales larger than $\simeq 30'$ foreground fluctuations are
essentially of Galactic origin and are minimum around
$60\,$GHz~\cite{Bennett et al. 2003b}. On smaller scales,
extragalactic sources take over, and the minimum shifts to
100--$150\,$GHz.

The minimum in the spectrum arises from a superposition of two
emission mechanisms. The radio emission is generally -- but not
always -- dominated by optically thin synchrotron, which decreases
with frequency with a power law spectrum (flux density $\propto
\nu^{\alpha}$, with $\alpha \sim -0.5$ to $-1.0$). At wavelengths
shorter than a few mm, again for typical sources, the emission is
dominated by thermal dust, whose spectrum increases sharply with
frequency ($\alpha \simeq 4$). Just because of such steep rise,
the frequency of the minimum is only weakly dependent on the
relative intensity of the two components. Moreover, the effect of
redshift on the dust emission peak is to some extent compensated
by the dust temperature increase associated with the luminosity
evolution of sources. Thus the smearing effect of the variance in
the source properties and of the redshift distribution is limited
and the minimum is quite deep.

Nevertheless, distinguishing the cosmological signal from
foreground contamination is one of the main challenges facing the
most sensitive CMB experiments. We will illustrate this issue with
reference to three recent exciting results: i) WMAP measurements
of the temperature-polarization correlation power
spectrum\cite{Bennett et al. 2003a,Kogut et al. 2003}; ii) the
detection of degree-scale polarization fluctuations by the DASI
experiment\cite{Kovac et al. 2002}; iii) the excess power on
arcminute scales detected by the CBI\cite{Mason et al.
2003,Pearson et al. 2003} and BIMA\cite{Dawson et al. 2002}
experiments.

\section{WMAP measurements of temperature-polarization
correlation}

Primordial adiabatic density fluctuations create, at decoupling,
polarization either perpendicular or parallel to the wave vector,
i.e. a curl-free pattern referred to as $E$-mode
polarization\cite{Hu White 1997,Zaldarriaga Seljak 1997}. Since
$E$-mode polarization is related to peculiar velocities on the
last scattering surface, the peaks of its power spectrum are
tightly related with those of temperature anisotropies, being out
of phase by $\pi/2$ since velocities are out of phase with density
perturbations like velocity and position of a harmonic oscillator:
velocities drop to zero at the compression or rarefaction maxima.

Given such tight correlation and the low level of CMB
polarization, compared to temperature ($T$) anisotropies, it is
easily understood that the $TE$ correlation has the largest
amplitude among CMB power spectra involving polarization. Its
detection is further eased by the fact that this is not true, in
general, for foregrounds. In the limiting case of a random
distribution of polarization angles, as in the case of
uncorrelated point sources, the power spectrum of the correlation
of $T$ with the Stokes parameters $Q$ and $U$, characterizing
linear polarization, vanishes:  $(T\{Q,U\})_{\rm
foreground}=\langle T^2p\cos(2\phi)\rangle=0$, where $p$ is the
polarization degree.

But in the case of Galactic synchrotron emission, the orderly
large scale magnetic field may translate in a $TE_{\rm Gal.
synch}$ comparable to, or larger than $TE_{\rm CMB}$. This is
illustrated by Fig.~1 where the power spectra at 30 GHz of the
$TE_{\rm Gal. synch}$ correlation predicted by two recent
models\cite{Baccigalupi et al. 2001,Giardino et al. 2002} are
compared with $TE_{\rm CMB}$. The $TE_{\rm Gal. synch}$ power
spectrum was computed in two cases: either as an average over the
full sky or restricting ourselves to high-galactic latitudes
($|b|>20^\circ$). $TE_{\rm Gal. synch}$ drops on small angular
scales (large multipoles $\ell$) due to the effect of the chaotic
component of the magnetic field. On the other hand, both models
predict, on large scales, a $TE_{\rm Gal. synch}$ power spectrum
exceeding that due to the CMB. Clearly, both models are highly
uncertain due to our poor knowledge of the Galactic polarized
emission, and the WMAP data already indicate that they grossly
overestimate $TE_{\rm Gal. synch}$. However, a significant
Galactic contribution to the $TE$ signal measured by WMAP at the
lowest frequencies cannot be ruled out easily.

\begin{figure}[t]
\centerline{\psfig{file=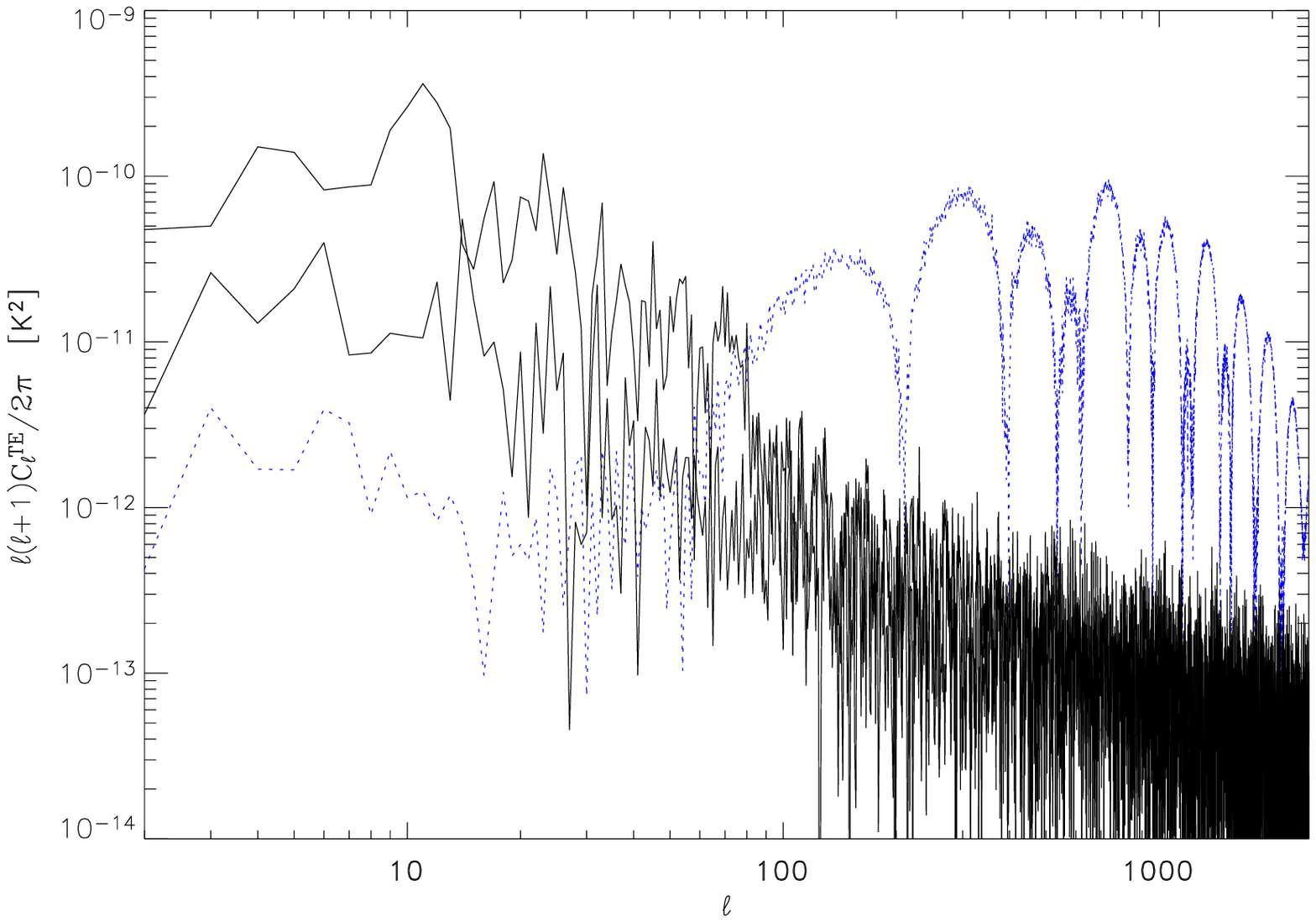,height=7cm,width=6.5cm}
\psfig{file=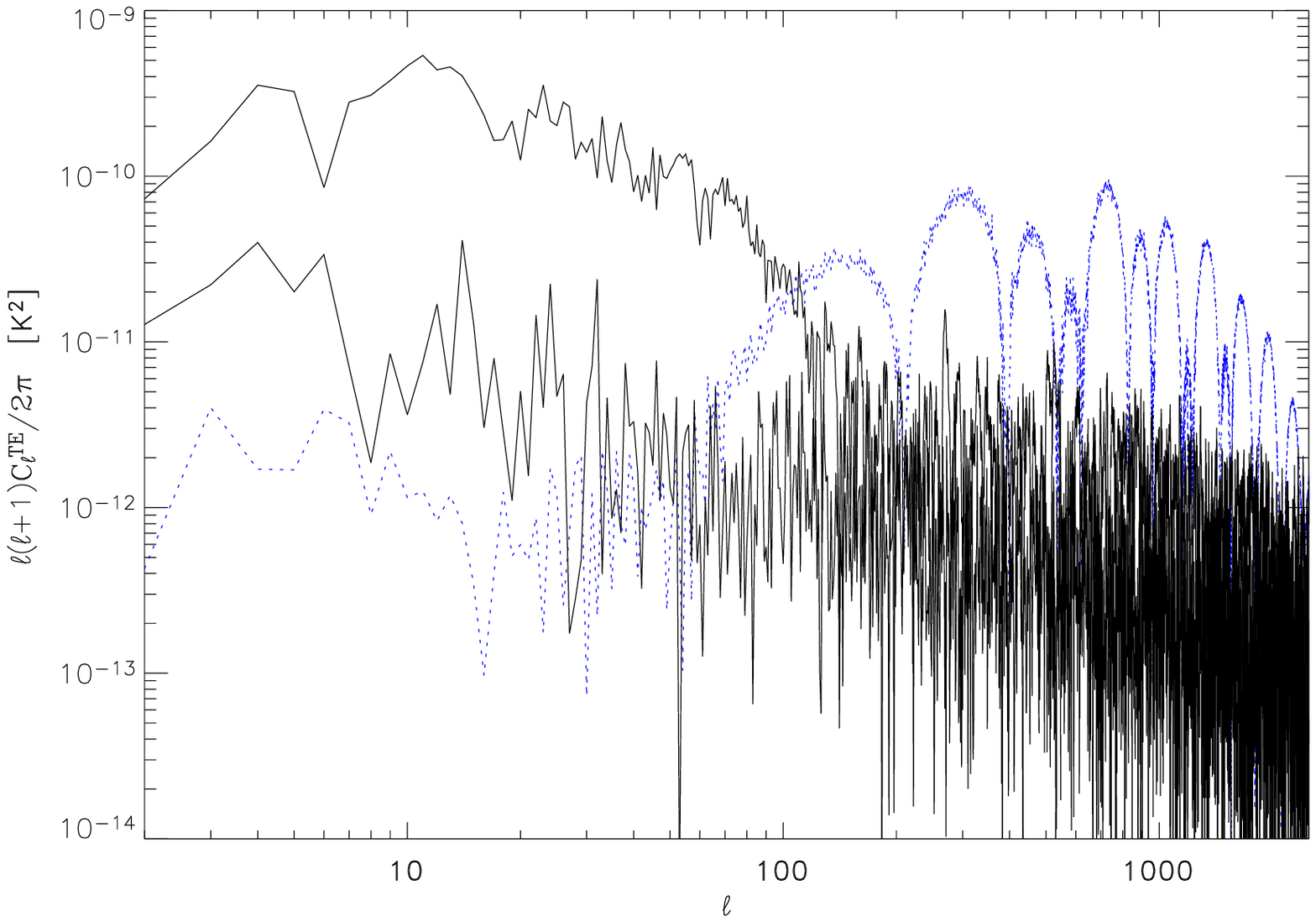,height=7cm,width=6.5cm }}
%\epsfxsize=12cm   %width of figure - will enlarge/reduce the figures
%\epsfbox{sorrento_cl_30_BACCI.ps,sorrento_cl_30_GIOVANNA.ps}
%\figurebox{2cm}{3cm}{} %to have a box alone
%\centerline{\epsfxsize=12cm\epsfbox{sorrento_cl_30_BACCI.ps,sorrento_cl_30_GIOVANNA.ps}}
\caption {Power spectrum of the $TE$ correlation for Galactic
synchrotron emission predicted by the models by\protect\cite{Baccigalupi
et al. 2001} (left panel) and \protect\cite{Giardino et al. 2002} (right),
extrapolated assuming that its antenna temperature has a spectral
index of $-2.9$ ($T_{\rm ant}\propto \nu^{-2.9}$). The upper curve
shows an all-sky average, the lower one refers to high Galactic
latitudes ($|b|>20^\circ$) only. Also shown is the CMB $TE$
spectrum (curve with multiple peaks, dominating at high
multipoles) based on a simulation using the best fit values of the
cosmological parameters obtained by \protect\cite{Spergel et al. 2003}
(last column of their Table 7) taking into account also large
scale structure data, plus a tensor component with a ratio of
tensor to scalar contributions to the temperature quadrupole
$T/S=0.3$. The re-ionization optical depth is $\tau=0.117$. }
\end{figure}

\begin{figure}[ht]
\centerline{\psfig{file=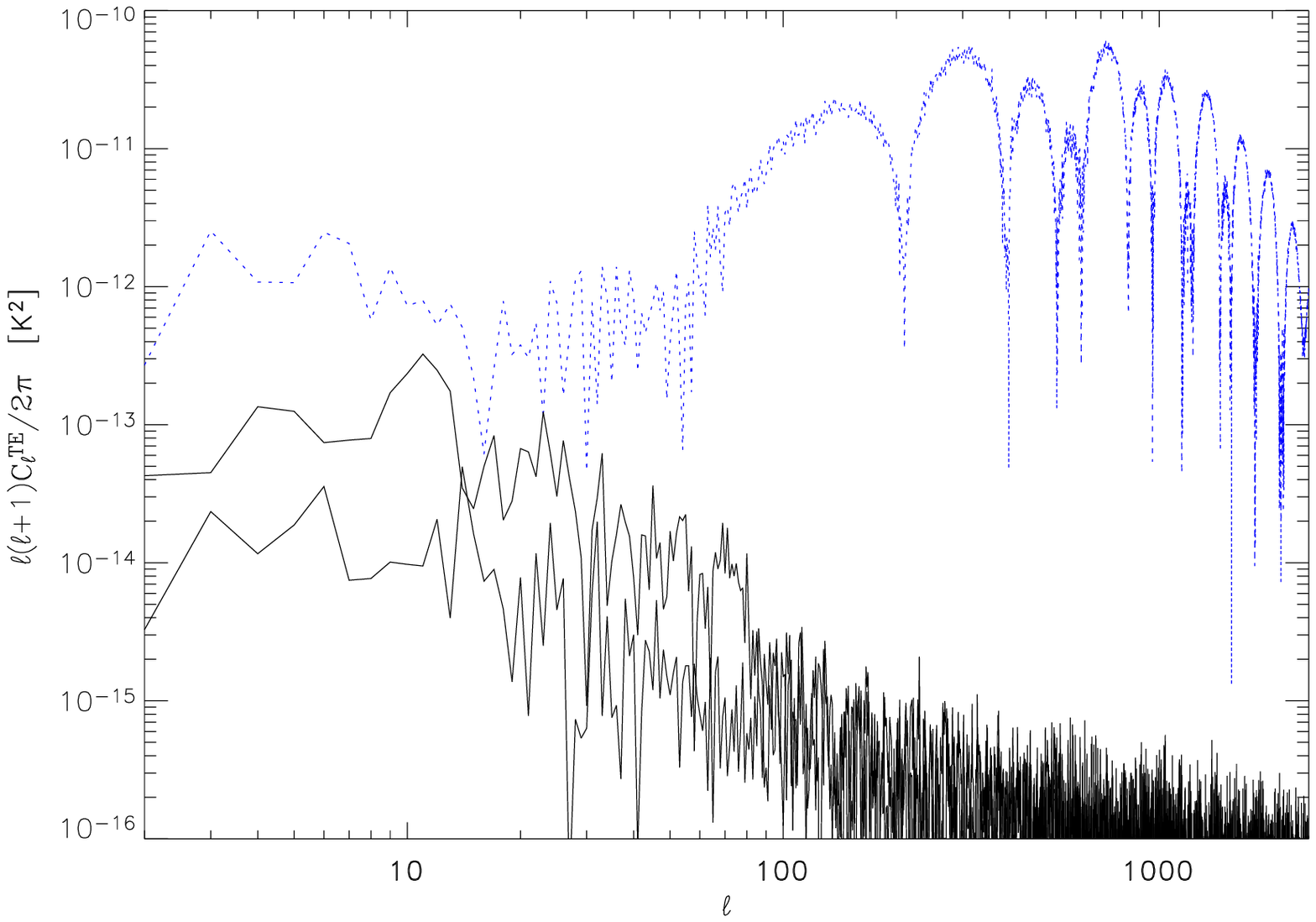,height=7cm,width=6.5cm}
\psfig{file=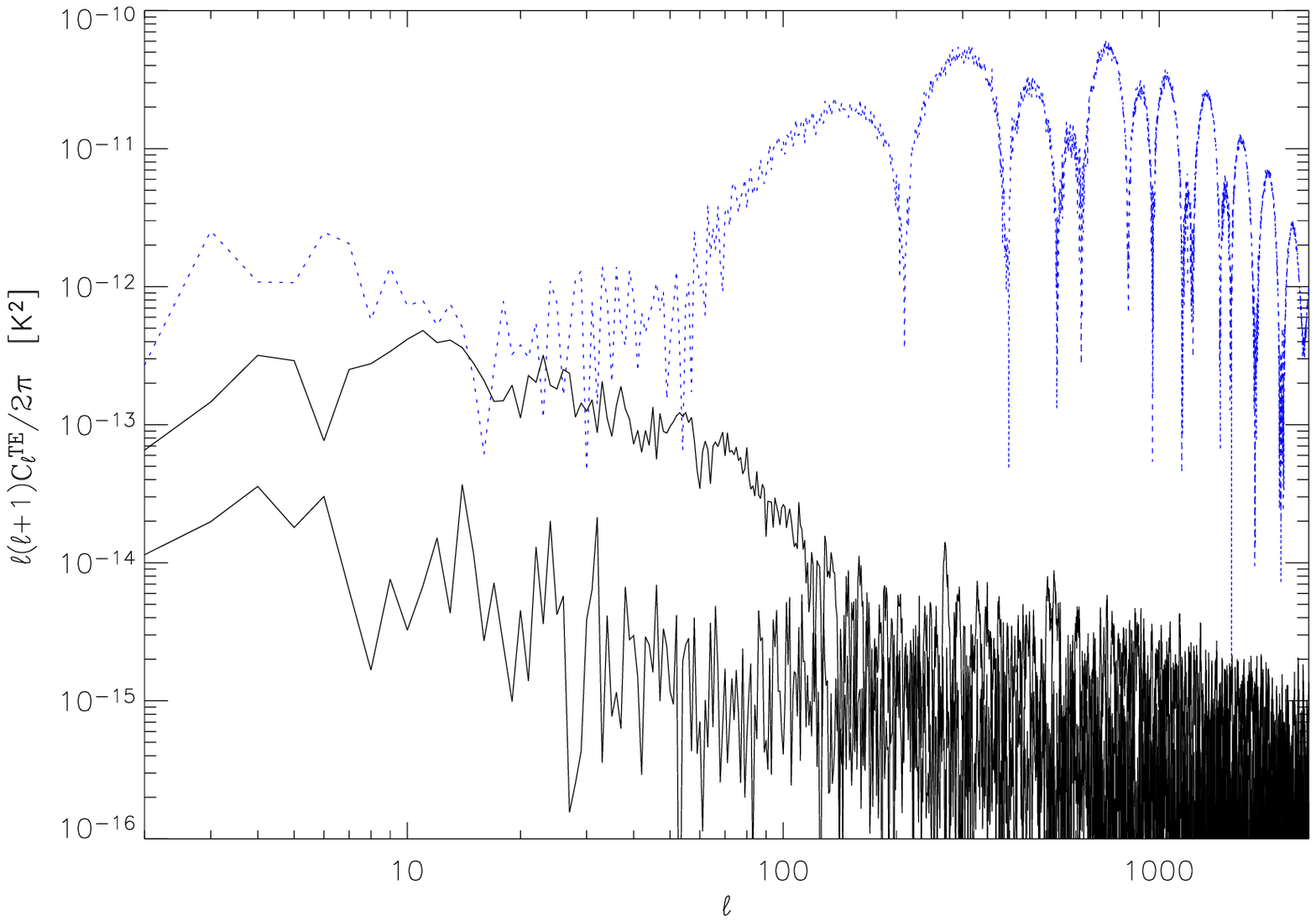,height=7cm,width=6.5cm }}
%\epsfxsize=12cm   %width of figure - will enlarge/reduce the figures
%\epsfbox{sorrento_cl_100_BACCI.ps,sorrento_cl_100_GIOVANNA.ps}
%\figurebox{2cm}{3cm}{} %to have a box alone
%\centerline{\epsfxsize=12cm\epsfbox{sorrento_cl_100_BACCI.ps,sorrento_cl_100_GIOVANNA.ps}}
\caption{Same as in Fig.~1, but a 100 GHz. }
%\end{center}
\end{figure}

Figure~2 shows that the synchrotron signal drops dramatically
(compared to the CMB signal) at high frequencies. This steep
frequency dependence backs up the conclusion that the $TE$ signal
detected by WMAP, which is consistent with being
frequency-independent\cite{Bennett et al. 2003a,Kogut et al.
2003}, is not seriously affected by Galactic contamination. Some
caution, however, may still be necessary in interpreting these
data. In fact, the recent ARCHEOPS experiment has
detected\cite{Benoit et al. 2003} large scale polarized emission
(polarization degree at the 4--5\% level) from Galactic dust
grains aligned by a coherent magnetic field coplanar to the
Galactic plane. Such emission, steeply rising with frequency, is
associated to the same magnetic field responsible for synchrotron
emission, and may to some extent compensate for the drop of the
synchrotron component at high frequencies.

\section{CMB polarization fluctuations on degree scales}

The first direct detection of CMB $E$-mode polarization may have
been achieved by the DASI experiment\cite{Kovac et al. 2002}, at
30 GHz, on degree scales. The most significant detection was
achieved at $\ell \sim 300$, which, at the frequency of
observations, roughly corresponds to the transition between the
large scales where the foreground power spectrum is expected to be
dominated by Galactic synchrotron and the smaller scales where the
most important contamination comes from polarized emission from
extragalactic radio sources.

The observational information on polarized synchrotron emission is
limited to low frequencies, which may be affected by Faraday
depolarization. On degree and sub-degree angular scales only a few
patches of the sky, mostly at low Galactic latitude have been
observed\cite{Duncan et al. 1997,Duncan et al. 1999,Uyaniker et
al. 1998,Uyaniker et al. 2003,Taylor et al. 2003}. In particular,
there are no published polarization maps in the region observed by
the DASI experiment\cite{Kovac et al. 2002}. Since the polarized
Galactic synchrotron emission varies substantially across the sky,
any estimate is highly uncertain. However, the estimates based on
the available information\cite{Tucci et al. 2000,Tucci et al.
2002,Baccigalupi et al. 2001,Baccigalupi et al. 2002,Burigana La
Porta 2002,Bruscoli et al. 2002,De Zotti 2002} do suggest that the
synchrotron contamination is likely to be below the signal
detected by the DASI experiment, consistent with the discussion
in\cite{Kovac et al. 2002}.

Estimates of polarization fluctuations due to extragalactic radio
sources are also affected by considerable uncertainty\cite{Sazhin
Korolev 1985,De Zotti et al. 1999,Mesa et al. 2002,Tucci et al.
2003}. Again, the dominant emission process is synchrotron, whose
radiation is intrinsically highly polarized. Yet the observed
polarization degree of most sources observed at MHz or GHz
frequencies is not higher than a few percent. Possible
explanations may be a strong Faraday depolarization or highly
turbulent magnetic fields. Thus, a strong frequency dependence of
the polarization degree may be expected. Measurements at high
frequencies are necessary!

A good opportunity in this direction is offered by the major
millimeter up-grade of the Australia Telescope Compact Array
(ATCA), which is going on since 2001.  In the commissioning phase,
the prototype receivers mounted on three antennas were
used\cite{Ricci et al. 2003} to take snapshots at 18.5 GHz of 250
out of the 258 Southern extragalactic sources in the complete 5
GHz 1 Jy sample\cite{Kuhr et al. 1981}. The configuration is very
compact, yielding a beam-width of $15.6''$ at the observing
frequency. Due to poor phase calibration, flux densities could
only be reliably determined for sources that can be successfully
fitted by the point-source model. This resulted in the rejection
of 53 sources. %A summary of measurements is given in
%Table~\ref{tab:kuhr}.

%\begin{table}
%\begin{center}
%\caption{Summary of polarization measurements of K\"uhr sources.
%\label{tab:kuhr}
%\begin{tabular}{|l|r|rr|rr|}
%\hline
%        &   All   & \multicolumn{2}{c|}{Flat}  &  \multicolumn{2}{c|}{Steep} \\
%\hline
%detections   & 170 & 114 & (80\%) & 56 & (49\%) \\
%upper limits &  27 &  12 & (8\%) & 15  & (13\%)  \\
%rejected     &  53 &  13 & (9\%) & 40 & (35\%)  \\
%not observed &   8 &   4 & (3\%) &  4 & (3\%)   \\
%\hline
%total        & 258 &  143 & & 115 & \\
%\hline
%\end{tabular}
%\end{center}
%\end{table}

\begin{figure}[t]
\centerline{\psfig{file=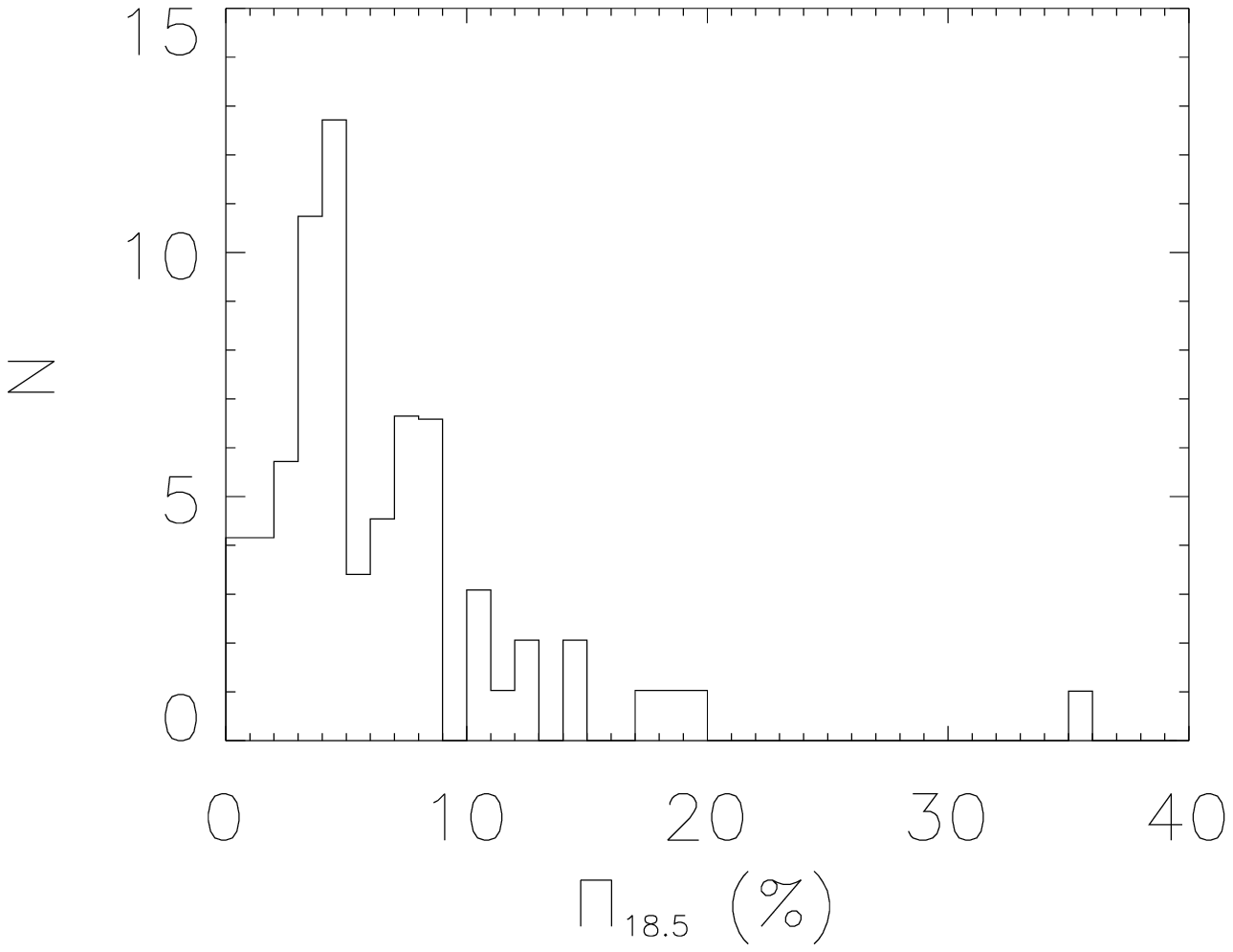,height=7cm,width=6.5cm}
\psfig{file=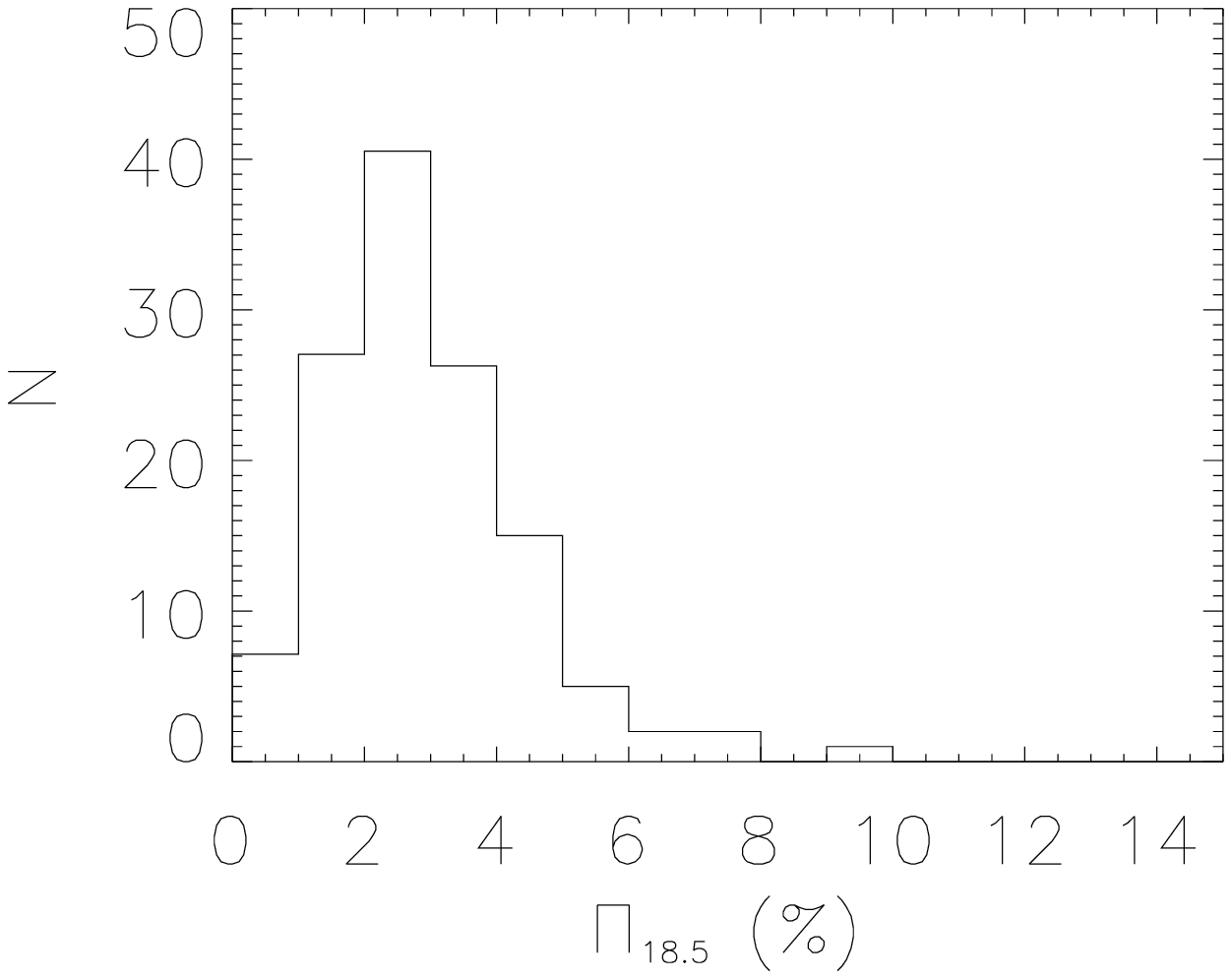,height=7cm,width=6.5cm }}
\caption{Distribution of the fractional polarization degree at
18.5 GHz of Southern K\"uhr steep- (left) and flat-spectrum
(right) sources \protect\cite{Ricci et al. 2003}.
%The morphological classification is from Stickel et al. (1994). }
}
\end{figure}

The majority of rejected sources (40 out of 53) are
steep-spectrum, consistent with the fact that steep-spectrum
sources are frequently extended. Also, these sources are obviously
fainter at high frequencies so that their polarization is more
difficult to measure. On the whole, polarization measurements were
obtained for only $\simeq 50\%$ of observed steep-spectrum
sources, and upper limits were determined for an additional 13\%.
Thus, the results for these sources are somewhat uncertain and
affected by a bias against extended sources. However, these
uncertainties are not critical for the estimate of polarization
fluctuations at the frequencies of CMB experiments, since in any
case, flat-spectrum sources dominate there.

The distribution of polarization degrees of steep-spectrum sources
at 18.5 GHz, $\Pi_{18.5}$, obtained using the Kaplan-Meier
estimator to take into account also upper limits, is shown in the
left-hand panel of Fig.~3. The median value is $\Pi_{18.5}\simeq
4.8\%$, and the mean is $6.5 \pm 0.7\%$. The 37 steep-spectrum
sources in common with the NVSS survey have a median polarization
degree at 1.4 GHz of only $\simeq 0.4\%$; there is thus clear
evidence of a strong frequency dependence of the polarization
degree for these sources.

The polarization degree was measured for $80\%$ of the
flat-spectrum sub-sample; including upper limits, we have data for
$88\%$ of these sources. The distribution of their polarization
degrees, obtained again with the Kaplan-Meier estimator, is shown
in the right-hand panel of Fig.~3. Allowing for upper limits, the
median value is found to be $\Pi_{18.5}\simeq 2.7\%$, the mean
$2.9 \pm 0.1\%$. The median polarization degree at 1.4 GHz of
sources in common with the NVSS survey is 1.4\%. Thus, there is a
much weaker frequency dependence of the polarization degree than
in the case of steep-spectrum sources. A closer look at the
distribution of $\Pi_{18.5}$ against $\Pi_{1.4}$ shows that
sources with $\Pi_{1.4}$ larger that $\simeq 2\%$ have, on
average, a similar polarization degree both at low and high
frequency, while sources with lower values of $\Pi_{1.4}$ have
relatively large $\Pi_{18.5}/\Pi_{1.4}$ ratios.

Armed with these results, we can now estimate the polarization
fluctuations due to extragalactic radio sources and combine them
with those due to Galactic synchrotron to derive the total
amplitude of foreground fluctuations. The results are shown in
Fig.~4. Uncertain as they are, these estimates are consistent with
a significant, but not dominant, foreground contamination of the
DASI results.

\begin{figure}[t]
\epsfxsize=12cm   %width of figure - will enlarge/reduce the figures
%\epsfbox{sorrento_DASI.ps}
%\figurebox{2cm}{3cm}{} %to have a box alone
\centerline{\epsfxsize=12cm\epsfysize=7cm\epsfbox{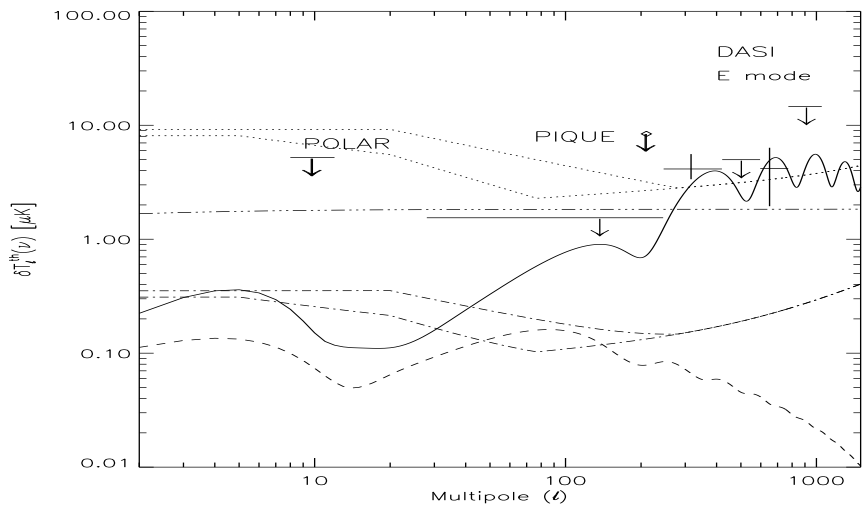}}
\caption{Estimated power spectra of Galactic synchrotron plus
point sources at 30 GHz (dotted line) and 100 GHz (dot-dashed
line) compared with the CMB $E$-mode (solid line) and $B$-mode
(dashed line) power spectra and with the results of the
PIQUE\protect\cite{Hedman et al. 2002}, POLAR \protect\cite{O'Dell et al. 2003},
and DASI\protect\cite{Kovac et al. 2002} experiments. The PIQUE experiment
operates at 90 GHz, while the POLAR and DASI experiment operate at
30 GHz.  The CMB power spectra refer to the same values of the
parameters as in Fig.~1. The horizontal line (dash-three dots)
shows a tentative estimate of the power spectrum of the polarized
emission from Galactic dust at 217 GHz, obtained from the average
power spectrum of dust emission at high Galactic latitudes
($|b|\ge 30^\circ$) determined by WMAP\protect\cite{Bennett et al. 2003b},
extrapolated to 217 GHz assuming a spectral slope of 2.2 in
antenna temperature ($T_{\rm ant} \propto \nu^{2.2}$), as
indicated by WMAP results, and adopting a polarization degree of
5\% \protect\cite{Benoit et al. 2003}. The pair of lines at 30 and 100
GHz, for $\ell < 200$, bound the range of synchrotron power
spectra found by Burigana \& La Porta\protect\cite{Burigana La Porta
2002}, extrapolated to 30 GHz assuming an antenna temperature
spectral index of $-2.9$. Note that the POLAR upper limit on large
angular scales is already below the lowest expectation for the
synchrotron signal. This may indicate that the synchrotron
spectral index is steeper than assumed (e.g. $\simeq -3.1$) or
that the polarized synchrotron emission is below average in area
covered by the POLAR experiment. The minimum at $\ell \sim
100$--200 in the foreground power spectra at 30 and 100 GHz
correspond to the transition between the angular scales where the
dominant polarized emission is synchrotron (lower values of
$\ell$) to the region where extragalactic radio sources take over.
}
\end{figure}

\section{The excess power on arcminute scales}

Evidences of statistically significant detections at 30 GHz of
arcminute scale fluctuations well in excess of predictions for
primordial anisotropies of the cosmic microwave background (CMB)
have recently been obtained by the CBI\cite{Mason et al. 2003} and
BIMA\cite{Dawson et al. 2002} experiments. The interpretation of
these results is still debated. Extragalactic sources are
potentially the dominant contributor to fluctuations on these
scales and must be carefully subtracted out. And indeed both
groups devoted a considerable effort for this purpose. However, as
discussed by~\cite{De Zotti et al. 2003}, a quite substantial
residual contribution to the CBI signal is difficult to rule out,
while the residual radio source contamination of the BIMA results
is likely to be very small.

If indeed the detected signal cannot be attributed to
extragalactic radio sources, its most likely source is the thermal
Sunyaev-Zeldovich\cite{Sunyaev Zeldovich 1972,Sunyaev Zeldovich
1980,Sunyaev Zeldovich 1981} (SZ) effect\cite{Gnedin Jaffe 2001}.
The SZ within rich clusters of galaxies has been extensively
investigated\cite{Komatsu Kitayama 1999,Bond et al. 2003}. The
estimated power spectrum was found to be very sensitive to the
normalization ($\sigma_8$) of the density perturbation spectrum. A
$\sigma_8\gsim 1$, somewhat higher than indicated by other data
sets, is apparently required to account for the CBI data.

\begin{figure}[ht]
\epsfxsize=12cm   %width of figure - will enlarge/reduce the figures
%\epsfbox{sorrento_DASI.ps}
%\figurebox{2cm}{3cm}{} %to have a box alone
\centerline{\epsfxsize=12cm\epsfysize=7cm\epsfbox{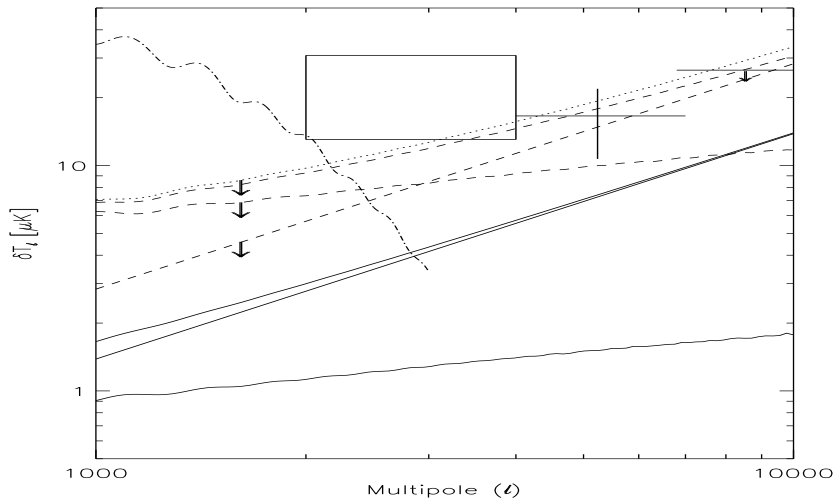}}
\caption{Possible contributions of galaxy-scale Sunyaev-Zeldovich
effects to the excess power on sub-arcmin scales detected by the
CBI\protect\cite{Mason et al. 2003} (box) and BIMA\protect\cite{Dawson et al.
2002} (cross and upper limit on the upper right corner)
experiments. The dot-dashed line shows the CMB power spectrum for
the set of cosmological parameters specified in the caption to
Fig.~1. The dashed lines refer to the contributions from
proto-galactic gas at virial temperature; the line with the
steepest slope shows the Poisson contribution, that with the
flattest slope shows the contribution due to clustering, while the
highest dashed line gives the sum of the the two contributions.
The downward pointing arrows indicate that the adopted values of
the parameters tend to maximize the signal. The solid lines show
the contributions due to the heating of the interstellar medium by
quasar feedback; again the line with the steepest and with the
flattest slope correspond to the Poisson and clustering
contributions, respectively, while the upper solid line shows the
sum of the two contributions. The dotted line is the sum of all
contributions.  }
\end{figure}

A preliminary analytical investigation of the SZ signals
associated to the formation of spheroidal galaxies has been
carried out by~\cite{De Zotti et al. 2003}. They point out that
the proto-galactic gas is expected to have a large thermal energy
content, leading to a detectable SZ signal, both when the
protogalaxy collapses with the gas shock-heated to the virial
temperature\cite{Rees Ostriker 1977,White  Rees 1978}, and in a
later phase as the result of strong feedback from a flaring active
nucleus\cite{Ikeuchi 1981,Natarajan et al. 1998,Natarajan
Sigurdssson 1999,Aghanim et al. 2000,Platania et al. 2002,Lapi et
al. 2003}.

These SZ signals, showing up on sub-arcmin scales, are potentially
able to account for the BIMA results. Proto-galactic gas heated at
the virial temperature and with a cooling time comparable with the
expansion time may provide the dominant contribution (see Fig.~5);
in this case we expect clustering fluctuations of amplitude
comparable to Poisson fluctuations, although with a flatter power
spectrum. As stressed by~\cite{De Zotti et al. 2003}, the estimate
of the signal from the gas at the virial temperature is based on
an over-simplified treatment. In fact, the thermal history of the
proto-galactic gas is quite complex: only a fraction of it may be
heated to the virial temperature\cite{Binney 2003,Birnboim Dekel
2003}; cooling may be relatively rapid in the densest regions
while significant heating may be provided by supernovae and by the
central active nucleus. On the other hand, the SZ effect turns out
to be an effective probe of the thermal state of the gas and of
its evolution so that its detection would provide unique
information on early phases of galaxy evolution, essentially
unaccessible by other means.

The SZ signals produced by strong feedback from the central active
nucleus have been discussed in the framework proposed
by\cite{Granato et al. 2001,Granato et al. 2003}, whereby the
interplay between star formation and nuclear activity in large
spheroidal galaxies is the key to overcome some of the crises of
the currently standard scenario for galaxy evolution.

This scenario envisages powerful quasar driven outflows, carrying
a considerable fraction of the quasar bolometric luminosity,
$L_{\rm bol}$, increasing as $L_{\rm bol}^{1/2}$ (for emission
close to the Eddington limit). For large enough galaxies ($M_{\rm
vir} {\gsim} 10^{12}\,M_\odot$) these outflows eventually sweep
out the interstellar medium (ISM) thus switching off the
star-formation on shorter timescales for more massive objects,
consistent with their observed $\alpha$-enhancement\cite{Thomas et
al. 2002}. The outflows are expected to be highly supersonic, and
therefore liable to induce strong shocks, transiently heating the
interstellar gas to temperatures exceeding the virial value by a
substantial factor. The associated SZ signals are correspondingly
stronger, albeit much rarer, than those due to the proto-galactic
gas heated at the virial temperature, discussed above.

A direct detection of individual SZ effects can be achieved by the
forthcoming large area high frequency surveys. We expect a surface
density of $\simeq 1\,\hbox{deg}^{-2}$ at $S_{30{\rm GHz}}\simeq
1\,$mJy. Of course, the counts at these relatively bright fluxes
are likely dominated by SZ effects in clusters of galaxies, which
however should be easily distinguishable because of their much
larger angular size and much lower redshift.

\medskip\noindent
Work supported in part by MIUR and ASI.

\end{document}